\documentclass[12pt,preprint]{aastex}
\usepackage{graphicx}

\shorttitle{Heating of active region cores}
\shortauthors{Durgesh Tripathi et al.}

\begin{document}

\title{Emission Measure Distribution and Heating of Two Active Region Cores}
\author{Durgesh Tripathi}
\affil{Inter-University Centre for Astronomy and Astrophysics, Post Bag 4,
Ganeshkhind, Pune 411 007, India}

\affil{Department of Applied Mathematics and Theoretical Physics, University of
Cambridge, Wilberforce Road, Cambridge CB3 0WA, UK}
\and

\author{James A. Klimchuk}
\affil{NASA Goddard Space Flight Center, Greenbelt, MD20771, USA }

\and

\author{Helen E. Mason}
\affil{Department of Applied Mathematics and Theoretical Physics, University of
Cambridge, Wilberforce Road, Cambridge CB3 0WA, UK}

\email{durgesh@iucaa.ernet.in}

\begin{abstract}
Using data from the Extreme-ultraviolet Imaging Spectrometer aboard
Hinode, we have studied the coronal plasma in the core of two active
regions. Concentrating on the area between opposite polarity moss,
we found emission measure distributions having an approximate
power-law form EM$\propto T^{2.4}$ from $\log\,T = 5.5$ up to a peak
at $\log\,T = 6.55$.  We show that the observations compare very
favourably with a simple model of nanoflare-heated loop strands.
They also appear to be consistent with more sophisticated nanoflare
models.  However, in the absence of additional constraints, steady
heating is also a viable explanation.
\end{abstract}


\section{Introduction}

It has proved very challenging to obtain definitive answers to the
very stubborn problem of how the solar corona is heated. The heating
of warm ($\sim 1$ MK) coronal loops is one exception. There is now
widespread agreement, though not complete consensus, that these are
multi-stranded structures heated impulsively by storms of nanoflares
(e.g., Klimchuk 2006, 2009 and references cited therein). Warm loops
account for only a fraction of the coronal plasma, however. The
corona also contains hot ($> 2$ MK) loops and diffuse emission at
all temperatures.  Much recent attention has focused on what is
often called the ''hot core" of active regions. This is the subject
of our study reported here.

A fundamental question concerns the temporal nature of coronal
heating. This is often discussed in terms of a dichotomy between
steady heating and nanoflares. What really matters for determining
the properties of the coronal plasma is the heating on individual
magnetic flux strands (field lines).  Few, if any, serious coronal
heating theories predict that the energy release is steady in this
regard \citep{klimchuk_2006}; rather, the heating occurs in
short-lived bursts. We refer to these small-scale impulsive bursts
as nanoflares. No particular physical mechanism is implied by the
term. It could be magnetic reconnection in tangled magnetic fields
as envisioned by \cite{parker_1983} and probably involving the
secondary instability of current sheets \citep{dahlburg_2005}; it
could be wave dissipation in drifting resonance layers
\citep{ofman_1998}; or it could be another mechanism altogether.

An important parameter is the time delay between successive nanoflares on a given
strand. By delay we mean the time interval between the end
of one nanoflare and the start of the next.  If the delay is much
shorter than a cooling time (tens to hundreds of seconds) then there
is minimal cooling between events and the heating is effectively
steady. We can approximate such a situation with perfectly steady
heating. It is becoming more common for the term ''steady heating"
to imply nanoflares that repeat with a high enough frequency that
the temperature hovers around one value and for ''nanoflare heating"
to imply nanoflares that repeat with a low enough frequency that the
plasma cools substantially. In addition, ''nanoflare heating"
usually implies heating that takes place in the corona. Impulsive
energy release in the chromosphere that is associated with spicules
\citep{depontieu_2011} is a different phenomenon, though it might
have similar physical attributes.

Attempts to determine whether active region cores have steady or
nanoflare heating have been inconclusive.  Some studies point out
that the intensities, electron densities, Doppler shifts, and
nonthermal broadening of observed emissions are often quite steady
in the sense that fluctuation amplitudes are small
\citep{antiochos,tripathi_moss_2010,brooks_warren}. A reasonable
conclusion is that the heating is steady.  However, this is only
certain if the cross-field spatial scale of the heating is
comparable to or larger than the resolution of the observations
(typically about 1000 km). There is good reason to believe that the
coronal magnetic field is structured on a much smaller scale
\citep{klimchuk_2006} and so it seems likely that the heating is
also structured on a much smaller scale. If so, direct evidence of
nanoflares would be washed out, since the emission from many
different strands would be averaged together. Such averaging occurs
both across the observational pixel and along the optically thin
line of sight. Hence, steady intensities, densities, Doppler shifts,
and nonthermal broadening are consistent with both steady heating
and nanoflare heating.  We note also that subtle variability in
seemingly steady intensities can provide good evidence for
nanoflares (Terzo et al. 2011; Viall \& Klimchuk 2011b).

Faced with the likelihood of unresolved structures, we must look to
other ways of distinguishing between steady and impulsive heating.
Investigating the distribution of plasma with temperature, as
quantified in the the emission measure distribution, EM($T$), is one
promising approach.  There have been many determinations of EM($T$)
through the years, but they have tended to average over large areas.
Results reported for active regions generally obey the power law
EM($T$) $\propto T^b$ up to a peak near 3 MK, where the index $b$
ranges between 1 and 3 (e.g., \citealt{dere_1982,dere_1993,brosius_1996}). 
When plotted on a
log-log scale, $b$ is the slope of a straight line.  Some authors
use the differential emission measure, DEM($T$), which is related to
the emission measure by EM($T$) = $T \times$ DEM($T$).  The slope of
DEM($T$) is smaller by an amount 1.0.

Observations that average over large areas tend to include both
coronal and transition region emissions.  Transition region here
refers not to a particular temperature range, but rather to the thin
region of steep temperature gradient at the base of the corona.  As
a rule of thumb, the temperature of the top of the transition region
is about 60\% of the maximum temperature in the strand
\citep{ebtel,bradshaw_2010}.  The transition region can therefore
reach very high temperatures depending on how hot the strand is.
Million degree moss seen in the 171 channels of SOHO/EIT, TRACE, and
SDO/AIA is just the transition region footpoints of hot core
strands. We use this later to determine the width of the core in the
active region we investigate here.  Moss can tell us a great deal
about the core plasma. For example, in a recent paper we showed that
the EM distribution of moss is better explained if the core is
heated impulsively than in a steady fashion
\citep{tripathi_moss_2010_apj}. Our results are compatible with
steady heating if the cross sectional area has the proper
temperature dependence in the transition region ''throat" of a
rapidly expanding strand, but we argued that this temperature
dependence is not likely to be maintained in the presence of spatial
and temporal pressure nonuniformities across the moss.

Two recent studies have explicitly avoided moss and concentrated on
the coronal emission in the cores of active regions.
\citet{warren_2011} studied a small ($10\times15$ arcsec$^2$)
inter-moss region between opposite magnetic polarities and found an
emission measure distribution that can be approximated by EM
$\propto T^{3.26}$ in the range $6.0 \le \log\,T \le 6.6$. The EM is
400 times smaller at $\log\,T = 5.8$ than it is at $\log\,T = 6.6$.
\citet{winebarger_2011} averaged over a $5\times25$ arcsec$^2$
inter-moss area from another active region and found a similar
power-law slope ($EM \propto T^{3.2}$) over the range $6.0 \le
\log\,T \le 6.5$. These distributions are considerably steeper than
what has been published in the past. Is this because the previous
studies averaged over a mixture of different features (core, moss,
non-core loops, etc.), or is it because the inter-moss regions of
the two new studies are, by chance, not typical of core plasma in
general?

To help answer this question, we have analyzed four inter-moss
regions within active region AR 10961 that was observed by multiple
spacecraft and one small inter-moss region
in active region AR 10980. In the next sections, we describe the observations and
our analysis of the data. We derive emission measure distributions
for the inter-moss regions, and by estimating the plasma in the
foreground of the core, we determine the EM distribution of the core
plasma itself. We then discuss the results in the context of related
measurements and the predictions of steady and nanoflare heating.

\section{Observations and Data Analysis} \label{obs}

\begin{figure}
\centering
\includegraphics[width=0.3\textwidth, angle=-90]{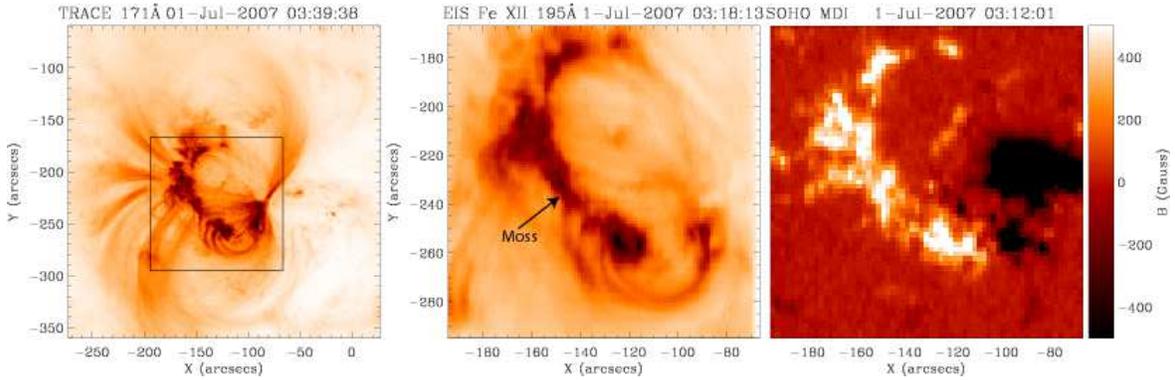}
\caption{Left panel: TRACE~171~{\AA} image showing the complete active
  region. The over-plotted box shows the EIS field of view. Middle
  panel: An EIS image obtained in \ion{Fe}{12}~195.12~{\AA}. The arrow
  identifies the moss. Right Panel: Co-aligned magnetogram
  obtained by the MDI instrument on SOHO. The left and middle panels
  are displayed in negative intensities. \label{context}}
\end{figure}

On July 01, 2007, the Extreme-ultraviolet Imaging Spectrometer
\citep[EIS;][]{eis} obtained a full spectral scan of the active
region AR 10961 with an exposure time of 25~sec using its 1{\arcsec}
slit. The field of view of the raster was 128{\arcsec} by
128{\arcsec} which basically covered most of the active region. The
EIS raster started at 03:18:13~UT and finished at 04:14:53~UT. A
full spectral scan of the active region was telemetered which
allowed us to choose spectral lines formed over a broad range of
temperature from $\log\, T = 5.5$ (\ion{Mg}{5}) to $\log\, T = 6.75$
(\ion{Fe}{17}, identified by \cite{fe_17} in EIS spectra).
Table~\ref{table:int} lists all the spectral lines used in this
study. The left panel in Fig.~\ref{context} shows the active region
recorded by the Transition Region And Coronal Explorer
\citep[TRACE;][]{trace} in the 171~{\AA} bandpass. The over-plotted
box marks the region which was rastered by EIS. An EIS image
obtained in \ion{Fe}{12}~$\lambda$195.12~{\AA} is shown in the
middle panel of Fig.~\ref{context}. The arrow identifies the moss.
The right panel shows a co-aligned magnetogram obtained by the MDI
instrument on SOHO. As can be seen in Fig.~\ref{context},  the
brightest moss regions show a strong correlation with the following
positive magnetic polarity \citep[cf.][]{tripathi_moss_2008,
noglik}. There also appears to be some  weak moss associated with
the leading negative magnetic polarity. We also note that most of
the core loops connect the brightest moss regions to the negative
polarity penumbra of the sunspot.

\begin{figure}
\centering
\includegraphics[width=0.5\textwidth]{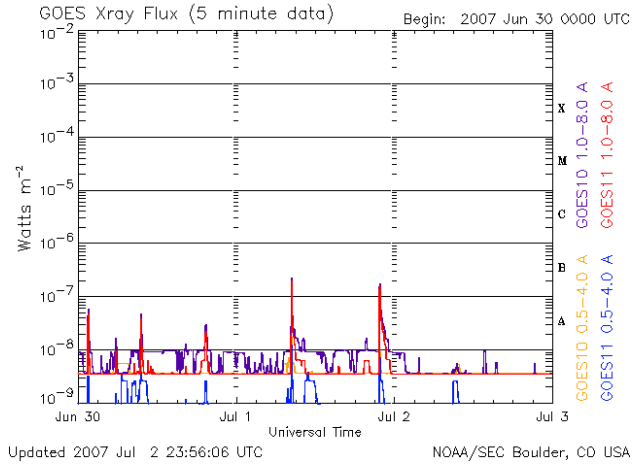}
\caption{GOES X-ray profile from June 30th to July 2nd 2007 showing
  minimal activity.\label{goes}}
\end{figure}

This active region raster is ideal for our study as it does not show any flaring activity
during the EIS raster. This is essential for our study because we
are mainly interested in the thermal structure of quiescent active
regions. Fig.~\ref{goes} displays the GOES X-ray flux for three days
from June 30, 2007. It is clear from the plot that the activity
remains minimal for three days except two small B-class flares
observed on July 1, 2007. However, both the flares occurred after
the EIS raster studies were completed.

The full spectral scan with a good exposure time allowed us to
choose spectral lines formed over a broad range of temperatures in
order to study the emission measure distribution of the plasma along
the line of sight (LOS) in the core the active region. For this
purpose we have selected relatively unblended lines (see
Table~\ref{table:int}) from the spectrum which covers a temperature
range of $\log T[\mathrm{K}]=5.5\mbox{--}6.8$. We have used the
standard EIS software provided in \textsl{SolarSoft} to process the
data and \textsl{eis\_auto\_fit}, also provided in
\textsl{SolarSoft}, for Gaussian fitting the spectral lines. For
de-blending some of the blended lines, we have used the same
procedure as in \citet{tripathi_moss_2010} based on the
recommendations of \cite{peter_artb, peter_dens}. For visual inspection,
we have also used data recorded by the TRACE and X-Ray Telescope \citep[XRT,][]{xrt}
on July 01, 2007 of the AR 10961. We have used standard software provided in
\textsl{SolarSoft} for processing these data.

\section{Visual Inspection of Hot and Warm Emission in the Core of the Active Region} \label{visual}

We have analysed the images recorded by TRACE and XRT in conjunction
with the images obtained from the EIS raster. Figure~\ref{trace_xrt}
displays co-aligned images from TRACE~171{\AA} (top left),
EIS~\ion{Fe}{12}~195{\AA} (top right), EIS~\ion{Fe}{15}~284{\AA}
(bottom left), and XRT using the Ti\_poly filter (bottom right). The
TRACE and XRT images were taken when the EIS slit was near the
middle of the raster. We used EIS \ion{Fe}{12}~195{\AA} and
\ion{Si}{10}~261{\AA} images to co-align the data taken by the two
different CCDs of EIS. The TRACE image was then co-aligned with EIS
\ion{Fe}{12} image and the XRT image was co-aligned with EIS
\ion{Fe}{15} image. We believe that the co-alignment achieved is quite accurate with an
error of about 3-5 arcsec. The images are displayed in negative intensities.

The figure clearly shows that both hot and warm emission exists in
the core of the active region. The emission consists of both a
diffuse component and distinct loops (labeled by an arrow in the
TRACE~171 image). It is not generally appreciated that the diffuse
component usually dominates the signal from active regions.
Identifiable loops are typically just modest enhancements on a
smoothly varying background \citep[e.g.,][]{delzanna_2003,
viall_2011a}. The loops are not as distinctive in the EIS images in
part because of the lower spatial resolution of the observations and
in part because hotter emission is inherently more fuzzy
\citep{tripathi_2009, guarrasi}. We note here that TRACE images
obtained before and after the EIS raster show that strong warm
emission is always present in the core of the active region.

\section{Emission Measure Distribution in Inter-moss Regions} \label{em_intermoss}

In order to determine the temperature distribution of the coronal
plasma within the core, we have performed an emission measure (EM)
analysis of three different inter-moss regions, labeled A ($\sim15\times10$ arcsec$^2$),
B ($\sim5\times5$ arcsec$^2$), and C ($\sim7\times7$ arcsec$^2$)
in Figure~\ref{masked_image}. The inter-moss regions are clearly separated
from both the bright moss seen in the TRACE~171 image and the strong
magnetic fields seen in the magnetogram.  They are
spaced along the axis of a magnetic arcade that spans the neutral
line, connecting opposite polarities. Regions B and C were chosen to
be small to be sure that their EM distributions reflect the
distribution of strands of different temperature (one temperature
per strand) and not the variation of temperature along each strand.
Because we sample three regions, rather than averaging over one
large area, we can be confident that our results are not strongly
biased by a small but exceptionally bright feature.

\begin{figure}
\centering
\includegraphics[width=0.6\textwidth]{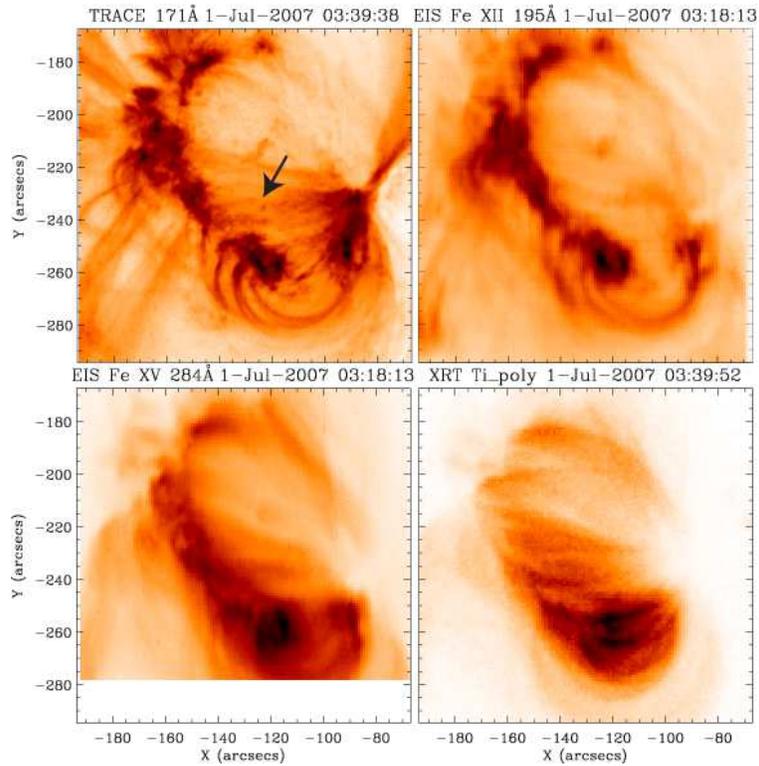}
\caption{Co-aligned TRACE~171{\AA} (top left), EIS
  \ion{Fe}{12}~195{\AA} (top right), EIS \ion{Fe}{15}~284{\AA} (bottom
  left) and XRT Ti\_ploy image shown in negative intensities taken on
  July 01, 2007. Images are displayed in negative colors. The arrow in
  the top left image labels the loop-like warm emission in the core of
  active regions. \label{trace_xrt}} \end{figure}

\begin{figure}
\centering
\includegraphics[width=0.4\textwidth]{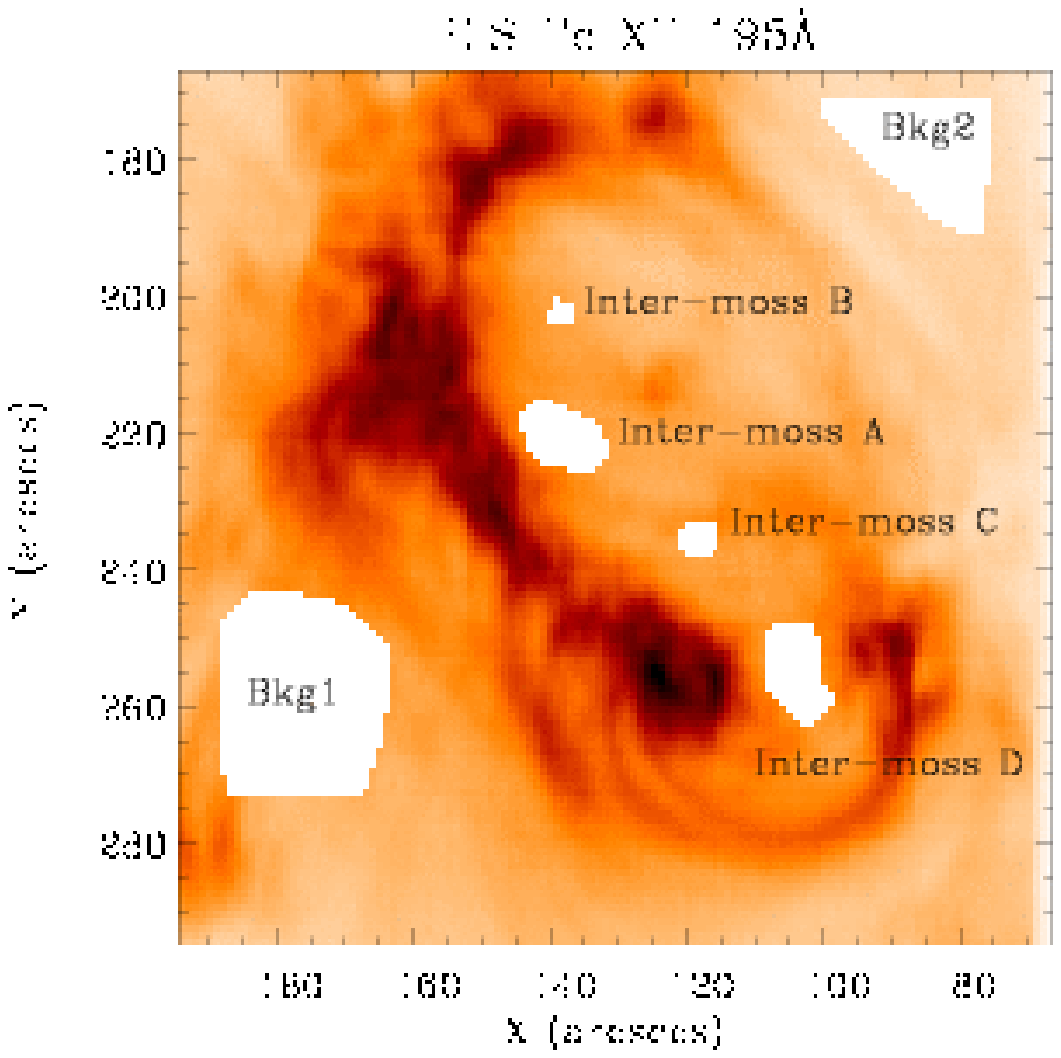}
\includegraphics[width=0.4\textwidth]{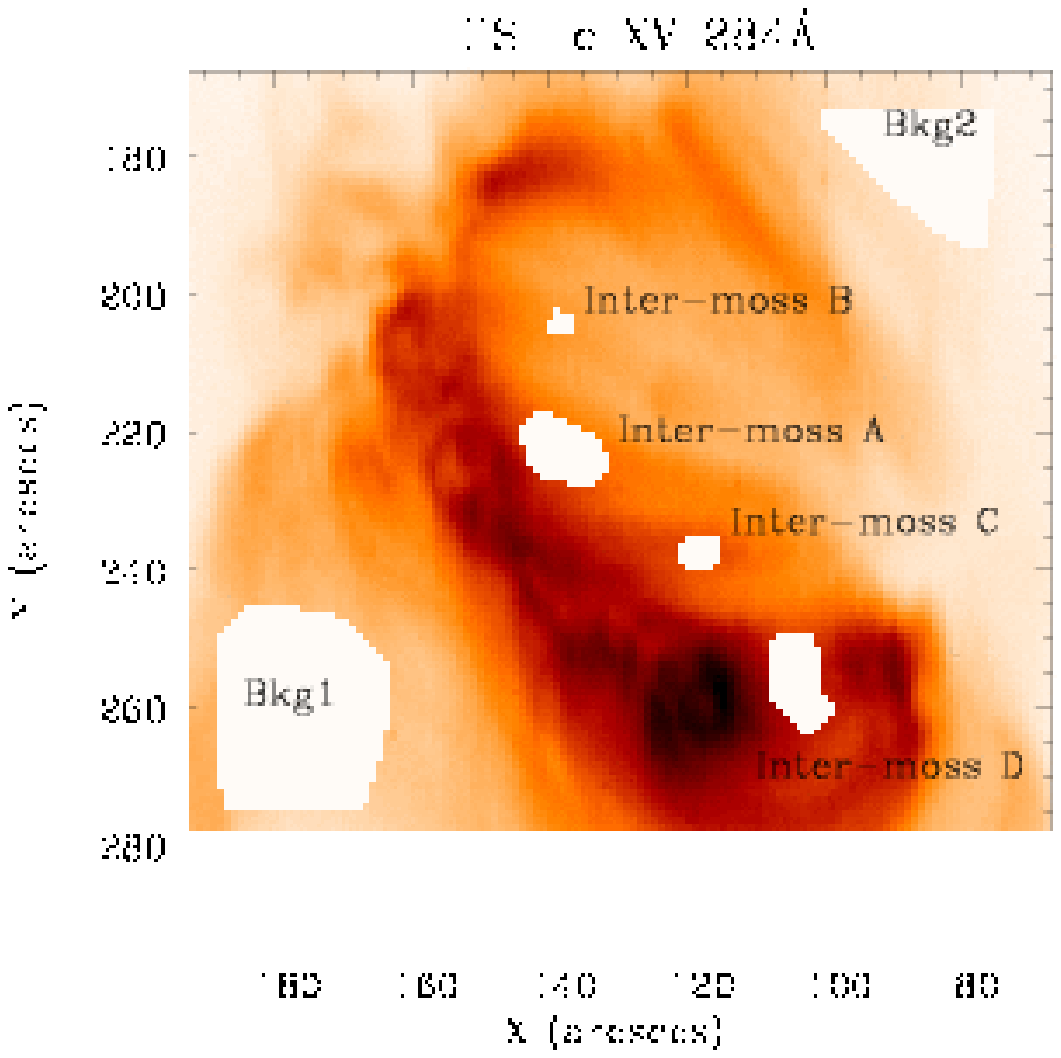}
\caption{\ion{Fe}{12}(left) and \ion{Fe}{15} images showing the four inter-moss regions,
  labelled as A, B, C and D and the background regions labelled as Bkg1
  and Bkg2 chosen for the study.}\label{masked_image}
\end{figure}

Although the three regions are all classified as inter-moss, they
appear quite different from each other in the images of
Fig.~\ref{trace_xrt}. The emission seems to become brighter with more
evidence of distinct loop structures in progressing from north to
south (i.e., from B to A to C). This could suggest substantially
different physical characteristics. However, we will show that the
plasma is in fact very similar in all three regions. This illustrates
how deceiving images can be when treated subjectively.  Meaningful
conclusions can only be drawn from a quantitative analysis of the
data. Table~\ref{table:int} presents the average intensities of
regions A, B, and C together with their 1-sigma errors
  obtained from Gaussian fitting for the 24 EIS spectral lines used in
our study. We also note that there is a systematic uncertainty of 22\%
  in intensities due to radiometric calibration as was
  measured by \cite{lang}.

We derive EM distributions from the intensities using the method
proposed by \cite{pottasch}.  For each spectral line, the EM at the
peak formation temperature, $T_{\rm max}$, is approximated by
assuming that the contribution function is constant and equal to its
average value over the temperature range $\log\,T_{\rm max}-0.15$ to
$\log\,T_{\rm max}+0.15$ and zero at all other temperatures. We
assume the ionization equilibrium values of \cite{chianti_v6} as given in Chianti~v.6.01
\citep{chianti_v1, chianti_v6}, and we consider both photospheric
\citep{photo_abund} and coronal \citep{coronal_abund} abundances.

\begin{figure}
\centering
\includegraphics[width=0.6\textwidth]{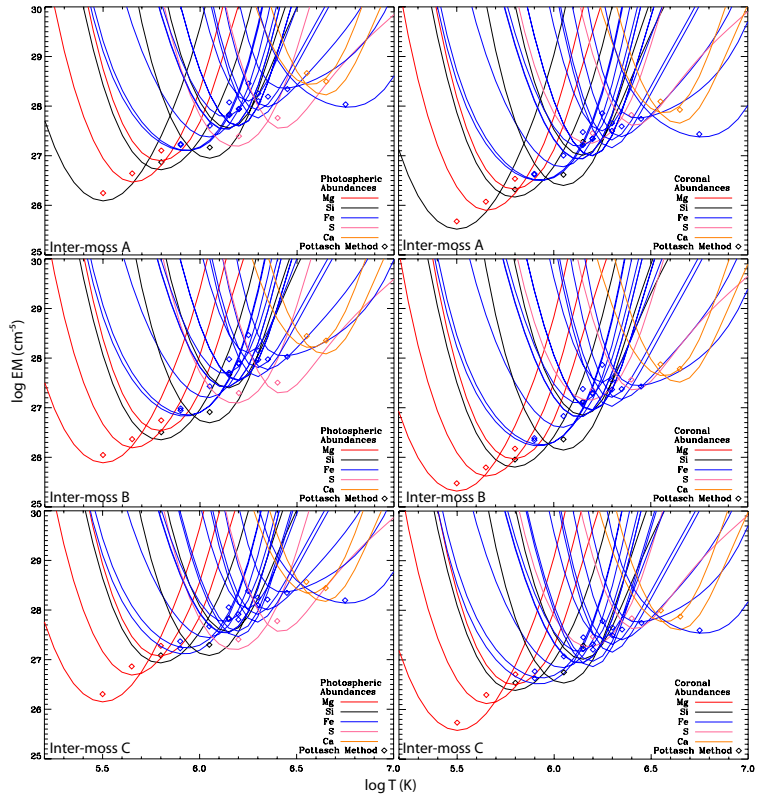}
\caption{EM loci plots (inverted 'U' curves) and emission measure
  derived using Pottasch method (diamonds) of each spectral lines for
  the three inter-moss regions labelled A, B and C in
  Fig.~\ref{masked_image}. The plot is color coded for different
  elements. \label{imoss_em_plot}}
\end{figure}

The resulting EM values are plotted as diamonds in
Figure~\ref{imoss_em_plot}.  The top, middle, and bottom panels are
for regions A, B, and C, respectively.  Panels on the left use
photospheric abundances and those on the right use coronal
abundances.  Coronal abundances provide more consistent results
(i.e., less scatter in the data), and so we conclude that they are
more appropriate.  \citet{warren_2011} reached the same conclusion
for the inter-moss region they studied.  Differences in the results
obtained with coronal and photospheric abundances are rather small,
however, as all lines with the exception of two sulphur lines come
from low-FIP (first ionization potential) elements. The figure is
color coded for the different elements. Sulphur is pink, and its two
lines fall at $\log\,T=6.20$ and $6.40$.  We will assume coronal
abundances for the remainder of the paper.  It is interesting to
note, however, that moss observations of this active region are more
consistent with photospheric abundances
\citep{tripathi_moss_2010_apj}.

\begin{figure}
\centering
\includegraphics[width=0.6\textwidth]{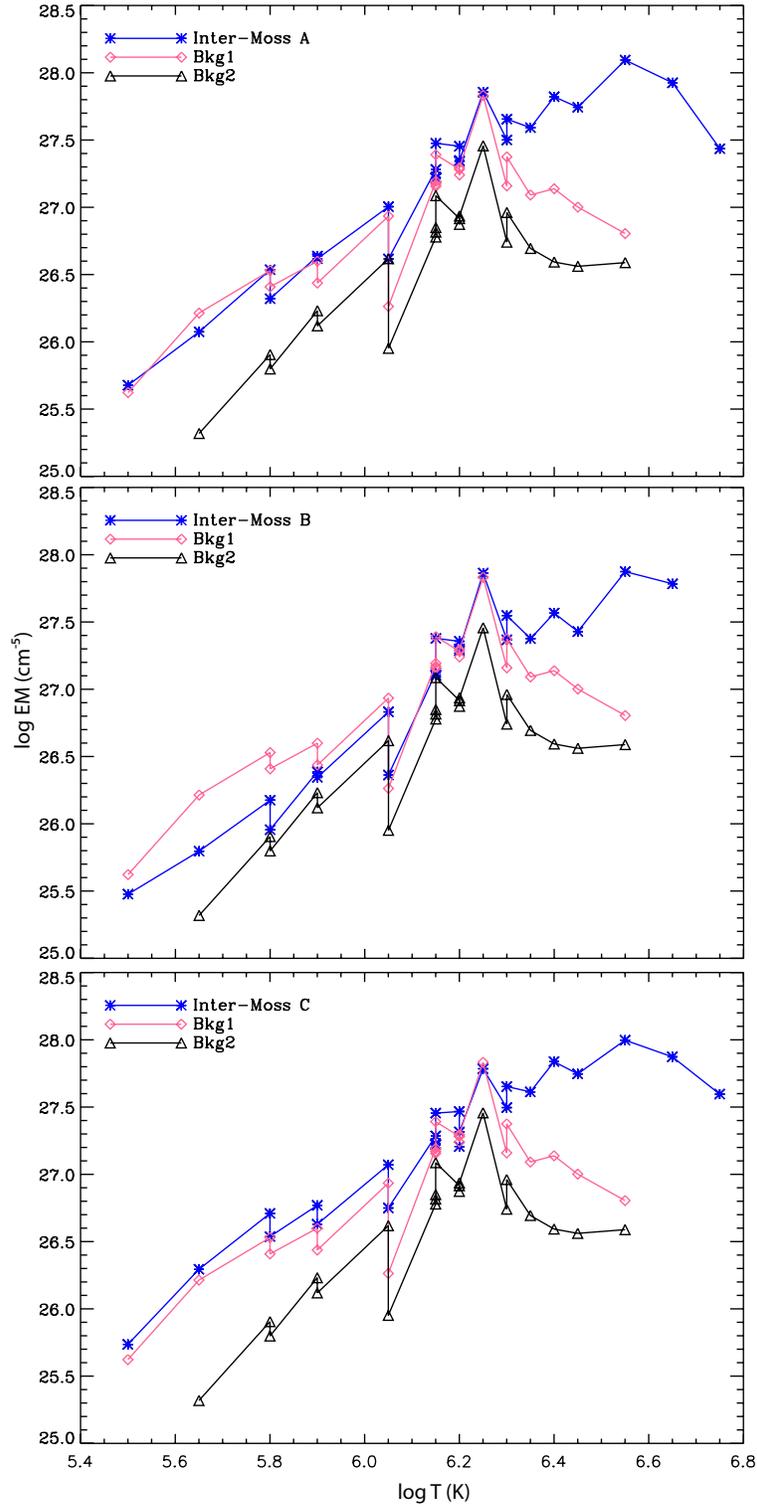}
\caption{Emission measure distribution in Inter-moss regions A, B and C
  over-plotted with the EMs of two background regions namely Bkg 1 and
  Bkg 2 labelled in Fig.~\ref{masked_image}. \label{imoss_bkg}}
\end{figure}

The inverted ''U" curves in the figure are EM loci plots
\citep[see][and references therein for details on EM loci]{em_loci}.  Each
curve gives the amount of emission measure that would be needed to
produce the observed line intensity if the plasma were isothermal at
the indicated temperature.  The curves have their minimum at the
temperature where the Pottasch value is plotted, as this is where
the line emits most efficiently and therefore requires the least
emission measure. As expected, the Pottasch values lie slightly
above the EM loci curves. It is clear that the Pottasch method gives
good results for the actual emission measure distribution that
produces the observed line intensities.  Were this not the case, the
Pottasch values would deviate significantly from the envelope of the
EM loci curves.  Henceforth, we use the terms EM distribution and EM
curve interchangeably to describe the connected Pottasch values.
They are shown in blue with asterisks in Figure~\ref{imoss_bkg}.

We estimate that the total uncertainty in the derived EM values is
approximately 50\%.  This includes errors resulting from photon
counting statistics, approximate atomic physics, uncertain elemental
abundances, and limitations in the Pottasch method.  The
corresponding error bars in Figure~\ref{imoss_bkg} are $\Delta \log$
EM $\approx \pm 0.2$.

The EM curves for all three inter-moss regions are very similar. To
within the uncertainties, they increase monotonically with
temperature from $\log\,T=5.5$ to a maximum at $\log\,T=6.55$. From
a simple linear fitting to this range we find that EM($T$) $\propto
T^b$ with slope $b = 2.33 \pm 0.15$, $2.47 \pm 0.22$, and $2.08 \pm
0.14$ for regions A, B, and C, respectively. The ratio of the EM at
$\log\,T=6.55$ to $\log\,T=5.8$ is 45, 62, and 23, where we have
used the actual values at these temperatures and not the linear fit.
Since we have two spectral lines at $\log\, T=5.8$, namely
\ion{Si}{7} and \ion{Mg}{7}, we have taken the average of two EM
values before obtaining the ratio. Although there is more plasma at
higher temperatures, it is clear that a substantial amount of plasma
is present at all temperatures along the LOS.

In addition to having similar slope, the EM curves for the three
inter-moss regions have similar magnitude. As already mentioned,
this contradicts the erroneous impression of greatly different
brightness that one gets from simply looking at the images in
Fig.~\ref{trace_xrt}. Such impressions have in the past led to the
incorrect notion that warm plasma exists primarily outside of the
core of active regions.  In fact, warm emission is usually stronger
in the core than outside \citep{viall_2011a}.

Not all of the line emission used to generate the inter-moss EM
curves in Figure~\ref{imoss_em_plot} comes from the core proper.
Some of it comes from foreground plasma that is part of the higher
arching field that overlies the core.  To estimate the contribution
from this foreground we generate EM curves for two ''background"
regions located outside of the moss, shown labeled as Bkg~1 and
Bkg~2 in Figure~\ref{masked_image}. The term background is something
of a misnomer in this case. As we discuss shortly, the EM of the
foreground to the core will be less than that of the background
regions.

The EM curves of the two background regions are shown repeated in
the three panels of Figure~\ref{imoss_bkg}.  Bkg~1 is indicated by
pink diamonds, and Bkg~2 is indicated by black triangles.  The
background curves have a similar shape to the inter-moss curves up
to $\log\,T=6.2$, and then they decrease at higher temperatures. The
magnitude is comparable to the inter-moss regions for Bkg~1 and
roughly a factor of 3 smaller for Bkg~2. The TRACE~171 image in
Figure 1 reveals that Bkg~1 contains distinctive fan loops that have
a southward trajectory and do not seem to overlie the core. For this
reason, we suggest that Bkg~2 is a more appropriate indicator of the
foreground plasma.

The EM of both background regions very likely over-estimates the EM
of the foreground plasma. Emission from the background regions is
integrated along the full line of sight from the solar surface to
the observer, whereas the foreground plasma begins at the top of the
core arcade. Since lower-lying plasma is denser and brighter due to
gravitational stratification, the difference in the integration is
substantial. We estimate the difference by considering a vertical
column of plasma with a density scale height $H_n$. It is easy to
show that the ratio of the integrated EM above height $z$ to the
integrated EM above the photosphere ($z = 0$) is

\begin{equation}
\frac{EM(z)}{EM(0)} = \exp \left(-\frac{2 z}{H_n}\right) .
\label{eq1}
\end{equation}

\noindent If we take $z$ to be the height of the core arcade,
equation \ref{eq1} is the amount by which the background EM must be
reduced to obtain a proper estimate of the foreground.  We estimate
from the moss in Fig.~\ref{trace_xrt} that the arcade width is
approximately 80 arcsec, or $6\times10^4$ km. Assuming that the
arcade is semi-circular, its height is then $3\times10^4$ km.
TRACE~171 emission is formed near $\log\,T=5.8$, for which the
gravitational scale height is $3\times10^4$ km. Taking this for
$H_n$, we obtain a reduction factor of 0.14 from equation \ref{eq1}.
This is only a lower limit, however, as distinguishable warm loops
are known to have a density scale height that is larger than
hydrostatic by up to a factor of 2 \citep{aschwanden01}. If the
diffuse component of the warm emission also behaves in this way, so
that $H_n = 6\times10^4$ km, then the reduction factor could be as
large as 0.4.

Summarizing, we conclude that the foreground plasma accounts for
between 5 and 40\% of the inter-moss emission measure at
$\log\,T=5.8$. The low extreme corresponds to Bkg~2 with a reduction
factor of 0.14, and the high extreme corresponds to Bkg~1 with a
reduction factor of 0.4. As discussed above, we believe that Bkg~2
is more representative of the foreground plasma, and therefore the
lower part of the range seems more likely.  We conclude that the
inter-moss EM curves of Figure~\ref{imoss_bkg} (blue asterisks) are
largely indicative of core plasma and not greatly contaminated by
foreground plasma. Figure~\ref{model} shows our best estimate of a
representative EM curve for the core itself.  The data points are
the averages of regions A, B, and C (linear averages, not
logarithmic) minus the averages of regions Bkg~1 and Bkg~2 reduced
by a factor of 0.25. The curve has a slope of approximately 2.4.

Although we believe our approach to estimating the foreground
emission is very sensible, we acknowledge that this is a very
difficult task and that the uncertainties may be large. Therefore,
in addition to what we take as the optimum foreground subtraction
described above (subtracting the average emission from Bkg1 and Bkg2
reduced by a factor of 0.25), we also consider two additional
possibilities. The first is no foreground subtraction, as presented
earlier. The second is a severe foreground subtraction, where we
subtract the full emission from region Bkg~2 without a reduction
factor.  The resulting best-fit slopes for the range $5.8 \le
\log\,T \le 6.55$ are given in Table~\ref{slope_5}. The slopes turn
out to be largely independent of how the foreground is estimated.

\begin{figure}
\centering
\includegraphics[width=0.45\textwidth]{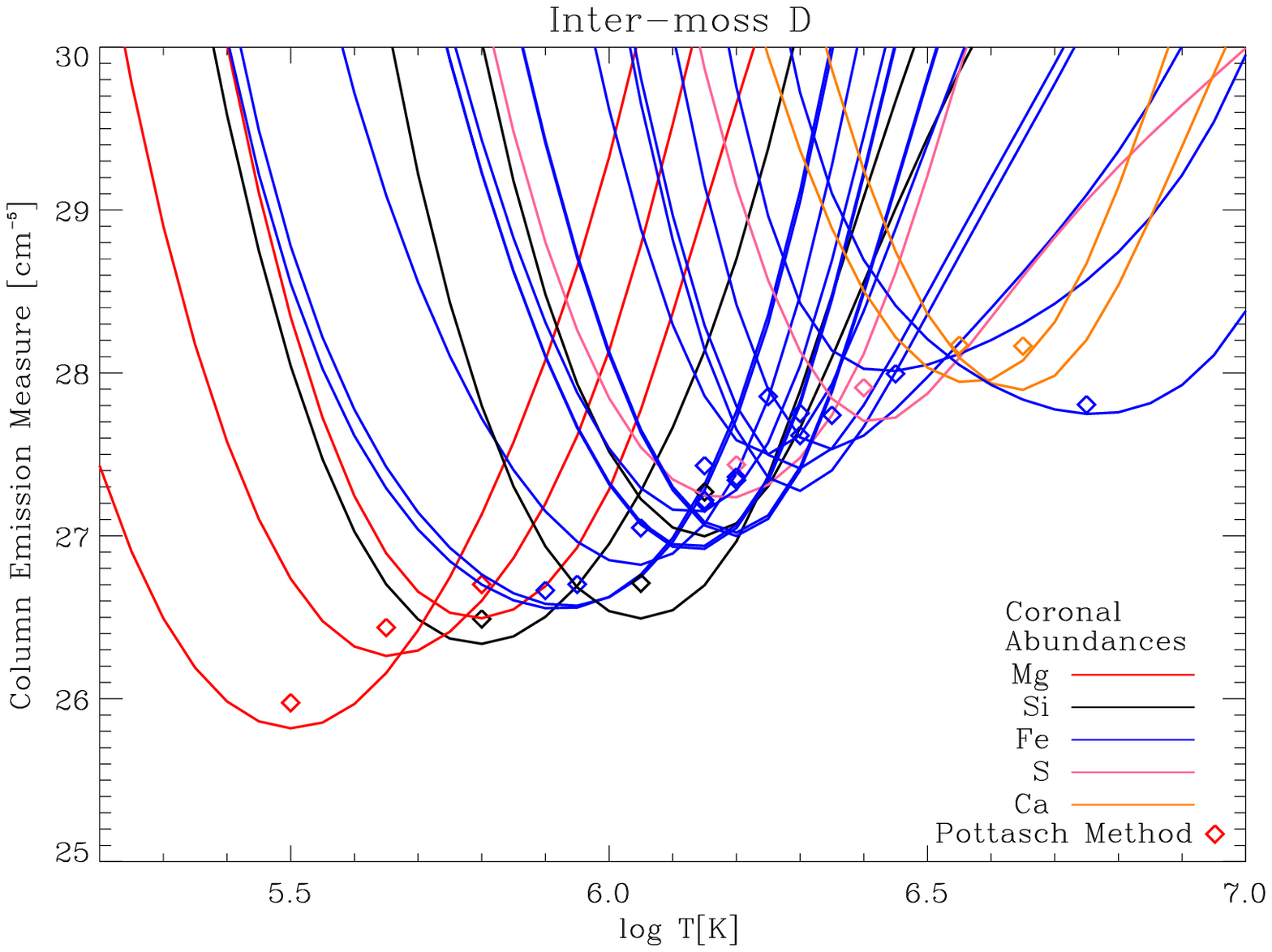}
\includegraphics[width=0.45\textwidth]{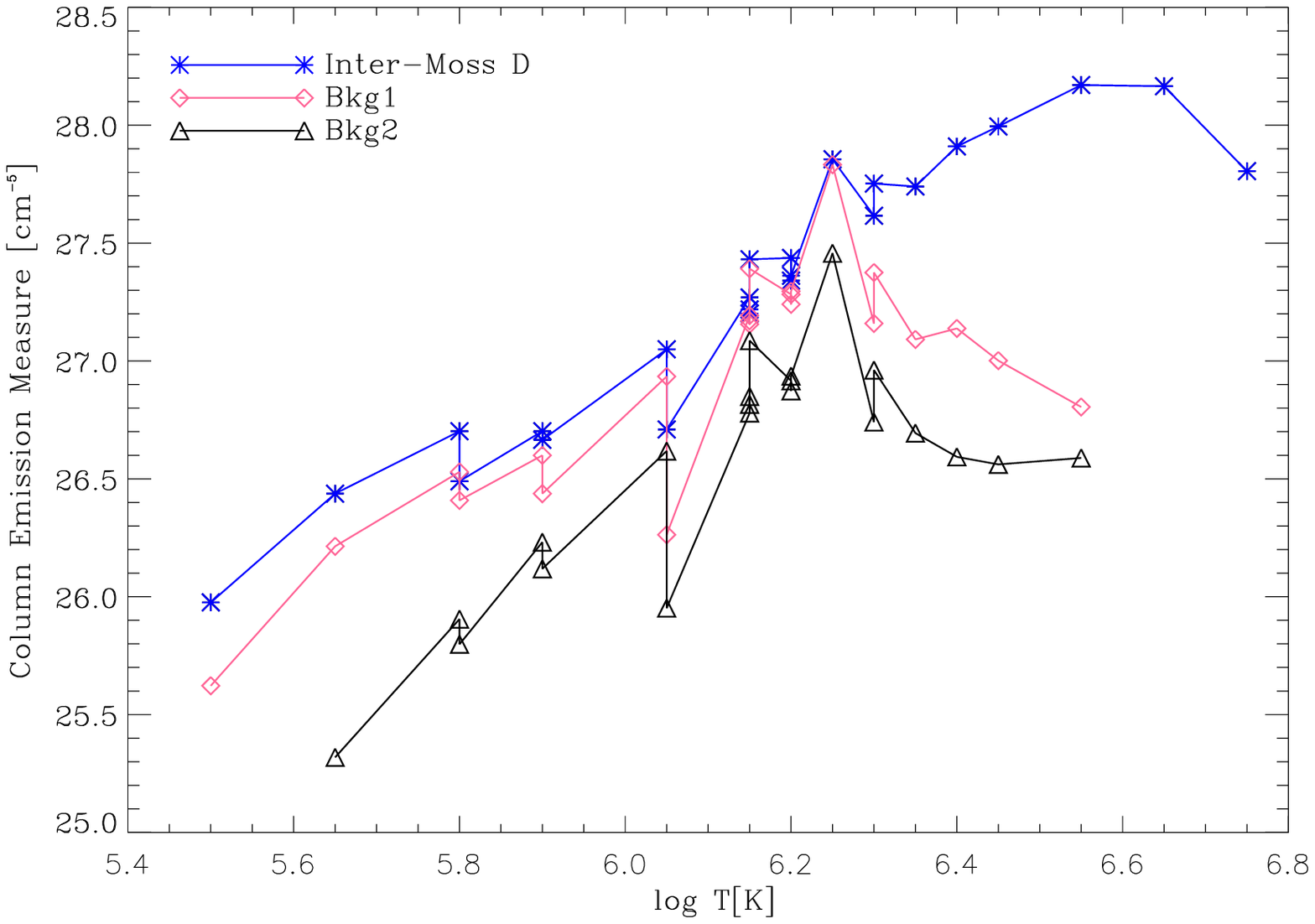}
\caption{Left panel: EM loci plots (inverted 'U' curves) and emission measure derived
using Pottasch method (diamonds) of each spectral line for inter-moss region D labelled
in Fig.~\ref{masked_image}. Right Panel:  Emission measure distribution in Inter-moss
region D over-plotted with the EMs of two background regions namely Bkg 1 and
Bkg 2 labelled in Fig.~\ref{masked_image}. \label{regionD}}
\end{figure}

We examined a fourth region, labeled D in Figure~\ref{masked_image}.
It includes the brightest XRT emission that does not directly
overlie strong photospheric field. We tentatively identify this as
an inter-moss region.  However, because the corridor between
opposite polarity strong field is so narrow ($<$ 10 arcsec), we
cannot be certain that there is no contamination from footpoint
emission. The EM loci curves and EM distribution for region D are
shown in Figure~\ref{regionD}.  They have a similar appearance to
those for regions A, B, and C.  The slopes obtained with the
different foreground subtractions are given in Table~\ref{slope_5}.
They are similar to the slopes for regions A, B, and C.  The
emission measure is 37 times smaller at $\log\,T=5.8$ than it is at
$\log\,T=6.55$ (with no foreground subtraction).

We studied a second active region, AR 10980, though in less detail.
It was observed by EIS on June 9, 2007 at 10:58:10~UT.
Figure~\ref{june9} shows the \ion{Fe}{12}~195.12 image (left panel)
and EM curves (right panel) obtained for the inter-moss region
($\sim 10\times10$~arcsec$^2$) labeled on the left. Again, coronal
abundances provide a more consistent result.  Without foreground
subtraction, the emission measure approximately follows a power law
of slope 2.33$\pm$0.19 up to a maximum at $\log\,T=6.55$. The EM is
69 times smaller at $\log\,T=5.8$ than it is at $\log\,T=6.55$.

The slopes given in Table~\ref{slope_5} are generally similar to
those reported previously for many other active regions, although
they are toward the steep end of the published range of 1--3 (e.g.,
\citealt{dere_1982,dere_1993,brosius_1996}). We must remember,
however, that those earlier studies may include moss and other
non-core plasma. \citet{brendan} studied an active region at the
limb and found very flat EM curves between $\log\,T = 6.0$ and
$6.5$. Those observations do not include moss, but they do include
non-core plasma along the horizontal line of sight.

A more relevant comparison is with the recent studies of
\citet{warren_2011} and \citet{winebarger_2011}, which also
concentrate on inter-moss regions. The considerably steeper slopes of 
3.26 and 3.2 found by \citet{warren_2011} and \citet{winebarger_2011} are 
from temperature ranges of $6.0 \le \log\,T \le 6.6$ and $6.0 \le \log\,T \le 6.5$,
respectively.  To make a more direct comparison with their results,
we have redone our analysis for the range $6.05 \le \log\,T \le 6.55$. 
Since Warren et al. subtracted no foreground before making their
measurements and Winebarger et al. subtracted only a weak foreground
(Winebarger 2011, private communication), the most appropriate
comparison is with "No Foreground".

The new slopes which we found for this temperature range vary from 1.75 to 3.05
(including estimated error) when no foreground emission is
subtracted. When foreground is subtracted, both optimum and severe, 
the values range from 1.95 to 3.15 (including estimated errors).  Region~D 
and AR10980 have the steepest slopes.  It should be noted that the values 
found in the other studies are steeper than the slopes we found in all cases,
indeed they are beyond the extreme limit of our values.

The fact that the slopes change depending on the width of the
temperature fitting range indicates that the assumption of a single
power law is not strictly valid.  Further analysis is required to
determine how the assumption fails.  Do broken power laws apply, and
if so, where are the breaks?  Is any range suitably approximated by
a power law?  We suggest that choosing the maximum of EM($T$) as the
high-temperature end of the fitting range may be dangerous, since
the distribution is likely to roll over near the maximum rather than
having a sharp peak.

We stress that we are confident in the basic validity of all three
studies discussed here (ours, Warren et al.'s, and Winebarger et
al.'s). The differences are real and indicate that not all active
region cores are alike.

\section{Discussion} \label{discussion}

The ultimate goal is to understand what these observations are
telling us about coronal heating.  Is the heating effectively steady
or does it take the form of low-frequency nanoflares with
substantial cooling between events?  A third possibility is that
plasma is heated in the chromosphere, not the corona, and ejected
upward as a spicule  \citep{depontieu_2011}, although see \citet{klimchuk_2011}.  We concentrate here on
the first two possibilities, since the spicule explanation is newly
developing, and many details are yet to be worked out.

If an observed inter-moss region is small enough that only short
segments of strands (loops) are observed, then any shape of emission
measure distribution is consistent with steady heating. Since the
strands are not evolving, the only requirement is to have the right
number of strands at each temperature.  We must then rely on other
information to determine whether this distribution of strands is
feasible. Fortunately, strands that are at or near static
equilibrium have well known and highly restrictive physical
constraints.  If the heating is not too asymmetric
\citep{winebarger_2002,patsourakos_etal_2004} and not too highly
concentrated near the footpoints \citep{klimchuk_2010}, then the
coronal temperature is uniquely related to the pressure and length
\citep{rosner_1978,craig_1978}. \citet{winebarger_2011} made use of
this in an impressive modeling effort that was a key part of their
study. From density sensitive line ratio measurements at the moss
footpoints and a potential magnetic field extrapolation, they
determined the pressures and lengths of the core strands that pass
through the LOS in the observed inter-moss region.  They then
determined the temperatures of the strands by assuming static
equilibrium and eventually obtained an EM distribution. The
distribution they found in this way is considerably more narrow than
what they observed. It matches the observations reasonably well in
the range $6.3 \le \log\,T \le 6.7$, but it drops abruptly to zero
at the extremes, unlike the observations. \cite{winebarger_2011}
suggest that the excess emission observed at warm temperatures is
due to foreground plasma above the core. They subtracted an estimate
of the foreground, but there is still much more warm EM in the
inter-moss region than the model predicts. It is possible that the
foreground was underestimated; on the other hand, no account was
made for gravitational stratification, which would tend to produce
an overestimate.  As mentioned earlier, treating the foreground is a
difficult business.

What about low-frequency nanoflares?  In a simple attempt to model
the core arcade of our active region, we performed a nanoflare
simulation using the EBTEL hydrodynamics code \citep{ebtel}.  We
considered one representative strand (loop) with a half length of
$2.4 \times 10^9$ cm.  For a semi-circular shape, this corresponds
to a radius of $1.5 \times 10^9$ cm, or half the estimated radius of
the arcade.  Thus, the strand has something of an average length for
all the strands in the arcade.  The nanoflares have a triangular
heating profile with a duration of 500 s and amplitude of 0.04 erg
cm$^{-3}$ s$^{-1}$. There is also a constant low-level background
heating of $10^{-6}$ erg cm$^{-3}$ s$^{-1}$. The nanoflares repeat
every 8000 s.  In order to obtain a predicted EM distribution for an
inter-moss observation, we assume that the time-averaged properties
of the coronal part of the model strand apply throughout a vertical
column of length $3 \times 10^9$ cm.  This is equal to the height of
the arcade. The time-averaged energy flux needed to maintain the
column is $3.75 \times 10^6$ erg cm$^{-2}$ s$^{-1}$, which is a very
reasonable value for active regions \citep{withbroe_1977}.

\begin{figure}
\centering
\includegraphics[width=0.8\textwidth]{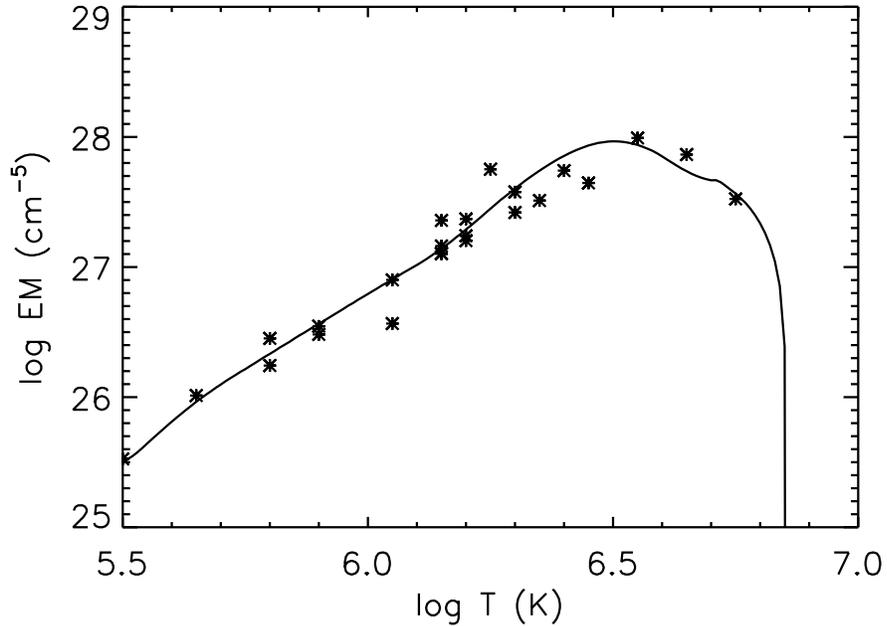}
\caption{Theoretical EM curve predicted by nanoflare heating model
  (solid line) and background subtracted and averaged observed
  emission measure (asterisks) for inter-moss A, B and C. See text for
  details. \label{model}}
\end{figure}

\begin{figure}
\centering
\includegraphics[width=0.8\textwidth]{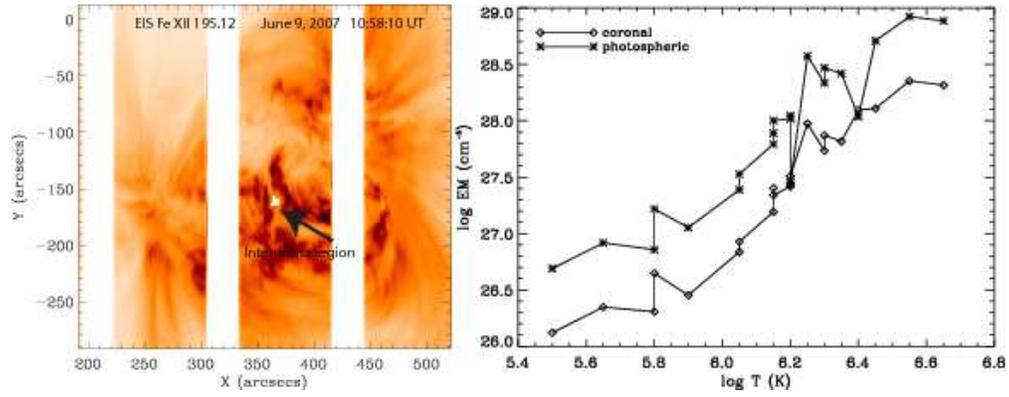}
\caption{Left panel: \ion{Fe}{12} image obtained from EIS raster on
  June 9, 2007. The arrow shows the inter-moss region chosen to study
  the EM distribution. Right panel: Emission measure distribution
  obtained for the inter-moss region shown in the left
  panel.\label{june9}}
\end{figure}

The solid curve in Fig.~\ref{model} shows the EM distribution from
our nanoflare model. The agreement with the observations is
excellent. Nearly all of the data points deviate less than
the $\Delta \log EM \approx \pm 0.2$ uncertainty. Using this EM
distribution we have also predicted intensities for different spectral lines
shown in Table~\ref{compare}. The predicted intensities are within 20-30\%
to the observed for most of the lines.  It is far too
premature, however, to conclude that core is necessarily heated by
nanoflares. Before any such conclusions can be drawn, we must
construct a more realistic and complete model, and we must consider
additional observational constraints, as was done by
\cite{winebarger_2011} for steady heating.  Such a modeling effort
is already underway. We also caution that the original EBTEL code is
most accurate during the heating and conductive cooling phases of
the strand evolution, and least accurate during the
radiation/enthalpy cooling phase. It tends to underestimate the
emission measure of the cooler plasma and therefore to overestimate
the slope of the EM distribution. We are in the process of making
improvements to the code \citep{cargill_2011}.

\citet{mulu-moore_2011} used the more sophisticated NRLFTM
hydrodynamics code to model nanoflare heating and found EM slopes
ranging from 2.0 to 2.3, depending on the loop length and nanoflare
energy (assuming coronal abundances). A composite model made by
combining the individual cases has a slope of 2.0.
 \citet{warren_2011} used the same code to predict a hot-to-warm
emission measure ratio similar to what we observe, but a factor of
10 smaller than what they observe.  Finally, Bradshaw \& Klimchuk
(2011, private communication) performed nanoflare simulations that
include the effects of non-equilibrium ionization and found slopes
consistent with those of \citet{mulu-moore_2011}.  They also found
that the slope increases with nanoflare duration and energy, as did
\citet{ebtel}.

As a final caveat, we note that the hottest plasma in a nanoflare
heated strand could deviate greatly from ionization equilibrium. The
intensities of emission lines such as the Fe XVII line in our study
may then be suppressed compared to what ionization equilibrium would
imply \citep{reale_2008, bradshaw_2011}. If that is case here, then
the actual emission measure at the hot extreme of our observed
distribution is greater than indicated in Fig.~\ref{model}. More
intense nanoflares would be needed to reproduce the observations.

In summary, we have studied the inter-moss cores of active regions
AR 10961 and AR 10980 and found  emission measure distributions with
a power law slope of approximately 2.4.  This is comparable to, but
on the steep end of what has been reported previously for active
regions as a whole.  However, it is considerably less steep than
what has been found recently in two other inter-moss regions
\citep[][]{warren_2011, winebarger_2011}. This raises an important
question as to what is typical. Future investigations of emission
measure distributions in the cores of many active regions are needed
before we have a clear answer. Our observations are in good
agreement with the predictions of a simple nanoflare model. However,
in the absence of additional observational constraints such as
pressure, plasma flows and non-thermal velocities, the observations
could equally well be explained by steady heating. Future studies
should include these additional constraints if possible.

\acknowledgments{DT and HEM acknowledge support from STFC. The work of
JAK was supported by the NASA Supporting Research and Technology and
Living With a Star Programs. CHIANTI is a collaborative project involving researchers
at NRL (USA) RAL (UK), and the Universities of: Cambridge (UK), George Mason (USA),
and Florence (Italy). We thank Dr Giulio Del Zanna for various discussions and Dr Peter
Young for providing his softwares to Solarsoft. We also thank Dr Amy Winebarger for providing constructive comments as a referee. Hinode is a Japanese mission developed
and launched by ISAS/JAXA, collaborating with NAOJ as a domestic partner, NASA and
STFC (UK) as international partners. Scientific operation of the Hinode mission is conducted
by the Hinode science team organized at ISAS/JAXA. This team mainly consists of scientists
from institutes in the partner countries. Support for the post-launch operation is provided by
JAXA and NAOJ (Japan), STFC (U.K.), NASA, ESA, and NSC (Norway). }


\begin{table}
\centering
\caption{Spectral lines used to study the emission measure distribution in inter-moss regions.
The intensities (I$_{obs}$) and 1-sigma errors (I$_{err}$) on the intensities are given for the
different regions. Intensity units are in ergs~cm$^{-2}$~s$^{-1}$~sr$^{-1}$. \label{table:int}}.

\begin{tabular}{lllrrrrrr}
\hline

Ion             &Wavelength     & log~T         & \multicolumn{2}{c}{Inter-moss A}      & \multicolumn{2}{c}{Inter-moss B}      & \multicolumn{2}{c}{Inter-moss C}\\
Name        &({\AA})            &(K)            &I$_{obs}$ & I$_{err}$                  & I$_{obs}$     &I$_{err}$              &I$_{obs}$          &I$_{err}$\\

\hline
\ion{Mg}{5}             &   276.58    & 5.50           & 9.5           &   0.8                         &  6.0          &   2.2                         &    10.9       &  1.5\\
\ion{Mg}{6}         &   268.99    & 5.65           &10.5           &   0.6                         &  5.5          &   1.9                         &    17.4       &  1.4\\
\ion{Si}{7}         &   275.36    & 5.80           & 62.0              &   1.2                         &  26.8             &   2.5                         &    102.3  &  3.0\\
\ion{Mg}{7}         &   278.40    & 5.80           & 88.9              &   1.7                         &  38.8             &  4.0                      &    135.9  &  4.0\\
\ion{Fe}{9}         &    197.86    & 5.90          &51.9           &   0.8                         &  29.1             &   1.9                         &    70.4       &  1.8\\
\ion{Fe}{9}         &   188.50    & 5.90           &112.3              &   1.7                         &  59.9             &   3.7                         &    115.3  &  3.3\\
\ion{Si}{9}         &    258.08    & 6.05           &31.9           &   1.3                         &  17.8             &   3.1                         &    43.3       &  3.5\\
\ion{Fe}{10}            &   184.54    & 6.05           &349.5              &   4.1                         &  234.8        &   9.9                         &    407.4  &  9.0\\
\ion{Fe}{11}            &   180.41    & 6.15           &1265.2         &   1.6                         &  963.9        &   39.4                        &    1282.7 &  29.7\\
\ion{Fe}{11}            &    188.23    & 6.15           &652.9              &   4.1                         &  510.5        &   10.8                        &    658.3  &  8.4\\
\ion{Fe}{11}            &       188.30    & 6.15           &454.42         &   3.8                         &  362.4        &   10.6                        &    432.9  &  7.4\\
\ion{Si}{10}        &       261.04    & 6.15           &136.2              &   2.0                         &  94.4             &   5.0                         &    137.5  &  4.0\\
\ion{Fe}{12}            &       192.39    & 6.20           &409.57         &   2.2                         &  362.6        &   5.9                         &    302.2  &  12.5\\
\ion{Fe}{12}            &       195.12    & 6.20           &1300.7         &   6.1                         &  1165.5       &   15.7                        &    1206.0 &  6.7\\
\ion{S}{10}         &       264.15    & 6.20           &97.5           &   1.6                         &  77.9             &   4.0                         &    100.5  &  3.0\\
\ion{Fe}{13}            &       202.04    & 6.25           &1012.7         &   6.3                         &  1029.1       &   16.9                        &    858.6  &  10.4\\
\ion{Fe}{14}            &       270.52    & 6.30           &421.3              &   2.8                         &  310.1        &   7.0                         &    416.9  &  5.5\\
\ion{Fe}{14}            &       274.20    & 6.30           &844.7              &   4.1                         &  653.1        &   10.4                        &    849.2  &  8.1\\
\ion{Fe}{15}            &   284.16    & 6.35           &6888.8         &   18.2                    &  4196.1       &   40.7                        &    7235.3 &  35.7\\
\ion{S}{13}         &       256.69    & 6.40           &611.9              &   4.9                         &  339.9        &   10.6                        &    635.2  &  9.5\\
\ion{Fe}{16}            &       262.98    & 6.45           &569.9              &   3.7                         &  276.5        &   7.6                         &    574.4  &  7.2\\
\ion{Ca}{14}        &  193.87    & 6.55           &138.7              &   1.3                         &  83.5             &   2.9                         &    110.8  &  2.2\\
\ion{Ca}{15}        &      200.97    & 6.65           &63.6           &   1.7                         &  45.9             &   3.9                         &    56.4       &  3.2\\
\ion{Fe}{17}        &      269.42    & 6.75           &7.4           &   0.8                         &  $-$             &   $-$                         &    10.7       &  1.8\\
\hline
\end{tabular}
\end{table}
\begin{table}
\centering
\caption{Measured slopes of the EM curves from $\log\,T=5.5$ to $\log\, T= 6.55$. The three columns from left to right
are for optimum foreground subtraction, no foreground and severe foreground subtraction respectively.\label{slope_5}}
\begin{tabular}{cccc}
\hline
Regions                 &  Optimum          & No foreground             & Severe \\
                            & Foreground            &                                   & Foreground\\
\hline
Inter-moss A            &   2.39$\pm$0.14       & 2.33$\pm$0.15             &   2.35$\pm$0.13    \\
Inter-moss B            &   2.64$\pm$0.22       & 2.47$\pm$0.22             &   2.70$\pm$0.25    \\
Inter-moss C            &   2.11$\pm$0.13       & 2.08$\pm$0.14             &   2.05$\pm$0.14    \\
Inter-moss D            &   2.17$\pm$0.15       & 2.14$\pm$0.14             &    2.13$\pm$0.17\\
AR10980             &       $--$                & 2.33$\pm$0.19                                  &$--$\\
\hline
\end{tabular}
\end{table}

\begin{table}
\centering
\caption{Comparison between observed and predicted intensities. Observed intensities are
the averages of intensities of regions A, B and C minus the averages of regions Bkg1 and Bkg2 reduced by a factor of 0.25.
Intensity units are in ergs~cm$^{-2}$~s$^{-1}$~sr$^{-1}$. \label{compare}}.
\begin{tabular}{lllrr}
\hline
Ion                         &Wavelength                 & log~T             & {Observed}         & {Predicted}      \\
Name                    &({\AA})                        &(K)                    &  {Intensity}               &{Intensity}                       \\
\hline
\ion{Mg}{5}                 &   276.58                  & 5.50              & 6.7                                &   7.3                    \\
\ion{Mg}{6}                 &   268.99                  & 5.65              & 7.3                        &   8.5                    \\
\ion{Si}{7}                 &   275.36                  & 5.80              & 42.4                               &   62.7                   \\
\ion{Mg}{7}                 &   278.40                  & 5.80              & 48.5                       &   41.2               \\
\ion{Fe}{9}                 &   197.86                  & 5.90                  & 36.0                       &   51.3                   \\
\ion{Fe}{9}                 &   188.50                  & 5.90              &72.8                        &  107.4                   \\
\ion{Si}{9}                 &   258.08                  & 6.05              &26.6                        & 59.7                 \\
\ion{Fe}{10}                &   184.54                  & 6.05              &238.2                       & 256.2                    \\
\ion{Fe}{11}                &   180.41                  & 6.15              &823.6                       & 909.4                    \\
\ion{Fe}{11}                &   188.23                  & 6.15              &492.8                               & 475.5                    \\
\ion{Fe}{11}                &   188.30                  & 6.15              &299.9                           & 181.73               \\
\ion{Si}{10}                &   261.04                  & 6.15              &89.6                        & 97.9                 \\
\ion{Fe}{12}                &   192.39                  & 6.20              &258.9                       &344.9                 \\
\ion{Fe}{12}                &   195.12                  & 6.20              &872.9                       &1072.3                    \\
\ion{S}{10}                 &   264.15                  & 6.20              &72.0                        & 64.9                 \\
\ion{Fe}{13}                &   202.04                  & 6.25              &676.4                       & 359.3                    \\
\ion{Fe}{14}                &   270.52                  & 6.30              &325.7                       & 479.6                    \\
\ion{Fe}{14}                &   274.20                  & 6.30              &636.8                       & 661.9                    \\
\ion{Fe}{15}                &   284.16                  & 6.35              &5452.9                      & 7680.6               \\
\ion{S}{13}                 &   256.69                  & 6.40              &492.8                       & 482.0                    \\
\ion{Fe}{16}                &   262.98                  & 6.45              &443.2                       &448.7                 \\
\ion{Ca}{14}                &  193.87                   & 6.55              &108.4                       &66.9                  \\
\ion{Ca}{15}                &   200.97                  & 6.65              &55.3                        &38.0                  \\
\ion{Fe}{17}                &   269.42                  & 6.75              &6.0                         &9.0                   \\
\hline
\end{tabular}
\end{table}


\end{document}